\documentstyle[psfig]{l-aa}

\input epsf 
\begin{document} 
\thesaurus{08.01.2; 08.02.4; 08.19.6; 08.22.3; 13.25.5}

\title{The ROSAT bright source RX~J0222.4+4729 : an active nearby 
short-period binary of the BY~Dra type 
\thanks{Based on observations
obtained at Observatoire de Haute-Provence (CNRS) with the 1.93--m, 1.20--m
and 0.80--m telescopes}}

\author{Claude Chevalier and Sergio A. Ilovaisky}
\institute{Observatoire de Haute-Provence (CNRS), F-04870 St.Michel
l'Observatoire, France}

\date{Received : 1996 December 4 / Accepted : 1997 March 28}

\maketitle 

\begin{abstract}

We report the discovery of a new BY~Dra-type binary identified as the
optical counterpart of the bright source RX~J0222.4+4729  detected
during the ROSAT All-Sky Survey (Voges et al. 1996). The star is a 
$V \sim 11.1$, near-by ($\sim 30$ pc), close spectroscopic binary with an
orbital period $P=0.46543 \pm 0.00001$ d. The absorption-line radial
velocities were obtained at
the 1.93-m Haute-Provence (OHP) telescope 
with the {\it Elodie} echelle spectrograph by on-line numerical
cross-correlation. 

The M0Ve primary exhibits strong Balmer and Ca~II H+K line emission,
placing this system amongst the most active BY~Dra stars. The width of
the cross-correlation function yields a projected rotational velocity
of $v \sin i \sim 85$ km/s. While only the primary contributes to the
continuum and the absorption line spectrum, the dM5e secondary is
detected through its H$\alpha$ emission. The mass ratio, estimated from
the amplitudes of the emission radial velocity curves, is $q =
M_{2}/M_{1} \sim 0.4$. 

CCD photometry in the $B$ and $V$ bands, obtained with the OHP 1.2--m
and 0.80--m telescopes, shows that the optical flux is modulated at the
spectroscopic period with a total amplitude of 0.2 mag and little or no color
change in $B-V$. The light curve, which can be attributed to
rotational modulation of the synchronized active primary star, shows extrema
near quadratures and also exhibits long-term variations in average
brightness (by 0.1 mag), which are accompanied by changes around the
photometric minimum. A secondary minimum appears at phase 0.5, indicating
a partial eclipse of the primary star.

In contrast with many other BY~Dra systems, the equivalent width of
the H$\alpha$ emission from the RX~J0222.4+4729 primary is directly
correlated with photospheric brightness, {\it i.e.} maxima and minima occur 
around the same phases in both curves. However, the minimum at mid-phase
in the  H$\alpha$ equivalent width is broader and deeper than the V-band
minimum at $\phi=0.5$ and appears shifted 
towards phase 0.45, suggesting that H$\alpha$ emission comes from extended
regions connecting the main starspot groups.

We find an X-ray to bolometric luminosity ratio of $\log (L_{x}/L_{bol})
\sim -3.1 \pm 0.14$ which supports the concept of saturation of coronal X-ray
emission for the most rapidly rotating late-type stars.

\keywords{stars: activity -- stars: starspots -- 
binaries: spectroscopic -- X-rays: stars} 
\end{abstract}

\section{Introduction} 
\markboth{C.Chevalier \& S.A.Ilovaisky : RX J0222.4+4729}
{C.Chevalier \& S.A.Ilovaisky : RX J0222.4+4729}

As part of an
identification program of X-ray sources discovered near the   
galactic plane during the ROSAT All-Sky Survey (Voges et al. 1996), we 
present here the results of the optical identification and follow-up
observations of the bright source RX~J0222.4+4729. The star
identified with  this source is shown to be an
active binary of the BY~Dra type with  one of the shortest (if not
the shortest) orbital periods.

\section{Optical identification} 
\subsection{Photometric observations}
CCD photometry was obtained at Haute-Provence
Observatory with the 0.8--m and 1.2--m telescopes. 
On the 0.8--m telescope we used the RCA No.3 CCD camera,
equipped with a thinned, back-illuminated 512$\times$323
chip with 30 $\mu$ pixels, yielding at the f/15 Cassegrain focus 
a 2.2$\times$3.5 arc--min field of view with a 
projected pixel size of 0.41 arc--sec. On the 1.2--m telescope
we used the TK512 No.2 CCD camera, equipped with a thinned, back-illuminated, 
AR-coated Tektronix 512$\times$512 chip with $27 \mu$ pixels, 
giving at the f/6 Newton focus a 6.5$\times$6.5 arc--min 
field of view with a 0.77 arc--sec projected pixel size.
Data were obtained on 1993 November, on 1994 November 
and December and on 1995 September and October. 
See Table~1 for the observing log. The CCD images were
reduced within the MIDAS environment following standard 
procedures. Small aperture photometry (allowing for pixels
lying partly within the aperture)
was used to perform relative measurements. The differential magnitudes
were computed with an aperture radius less than or equal to the
stellar FWHM. On frames used for calibration a growth-curve method
(Howell 1989, Stetson 1990) was applied to stars without close
neighbours in their profile wings, yielding the average correction needed
to convert the small aperture measurements into magnitudes
integrated over the entire stellar profile.

\begin{figure}
\epsfxsize=8.8cm
\epsfbox{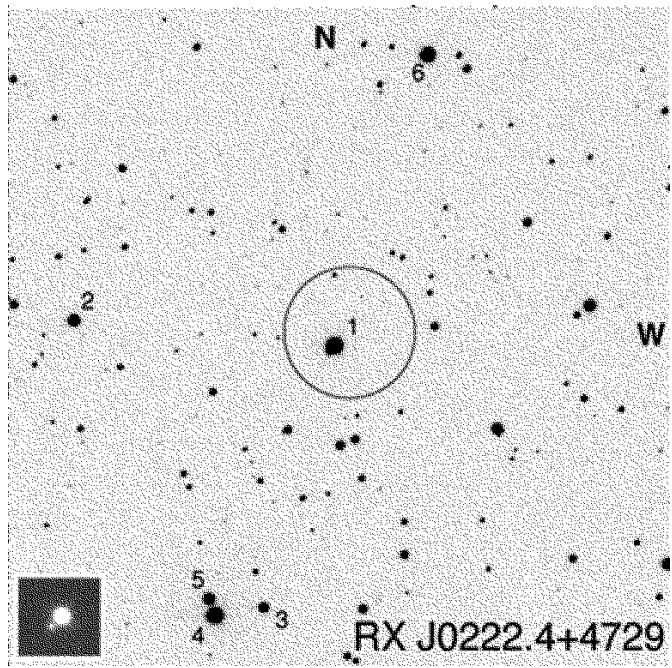}
\caption{Finding chart for RX~J0222.4+4729 made from a $V$--band CCD image taken
at the OHP 1.2--m telescope. The preliminary 90\% confidence error circle of 40" radius
is shown centered at the ROSAT X-ray position. The bright star within the circle
is the proposed $V=11$ optical counterpart, called Star~1. Stars~2 to 6 were used as
comparison and check stars. The magnitudes and colors of the numbered stars
are given in Table~2. The inset frame at the lower left is a close-up view
of the area around Star 1 obtained from a CCD frame taken at the 0.8--m telescope
and displayed at twice the scale to show the $V=15.1$ neighbor at 4.5" SE of
Star~1.}
\end{figure}

Figure 1 is a finding chart for the RX~J0222.4+4729 field
made from a 1.2--m CCD frame.
The 90\% confidence 40" radius ROSAT error box of RX~J0222.4+4729 
contains an 11th magnitude star  (GSC 3298.01172, Star~1 in Figure 1) lying
about 13" SE from the X-ray position given in the ROSAT 
All-Sky Bright Source Catalog (Voges et al. 1996) where the total positional 
error is listed as $\pm$8". 
The optical astrometric position for Star~1 is :
$\alpha_{2000} =$ 02h 22m 25.98s, $\delta_{2000} = +47^{\circ}$ 29' 17.9"
as derived from the Digitized Sky Survey CD-ROM set. This
position should be accurate to $\pm0.6$" r.m.s. 
(V\'eron-Cetty and V\'eron 1996).

\begin{table}
\caption{Photometry Log} 
\begin{tabular}[t]{ccp{1cm}c}
\hline
Date & Telescope & No. of Exp.  &  Filter \\
\hline

1993 Nov 11-12  &  0.8m OHP & 43 &  V \\
1993 Nov 12-13  &    "      & 70 & " \\
1993 Nov 13-14  &    "      & 39 & " \\
1993 Nov 14-15  &    "      & 55 & " \\
1993 Nov 16-17  &    "      & 44 & " \\
1993 Nov 17-18  &    "      & 35 & " \\   
1993 Nov 19-20  &    "      & 15 & " \\
1993 Nov 20-21  &    "      & 29 &  " \\
1993 Nov 21-22  &    "      & 19 &  " \\
 & & & \\
1994 Nov 15-16  &  1.2m OHP & 50 &  V \\
1994 Nov 25-26  &    "      & 27 &  " \\
1994 Dec 11-12  &    "      & 16 &  " \\
1994 Dec 13-14  &    "      & 46 &  " \\
1994 Dec 14-15  &    "      & 46 &  " \\
 & & & \\
1995 Sep 13-14  &  1.2m OHP & 17 & B,V,R,I \\
1995 Sep 14-15  &    "      & 26 &  B,V \\
1995 Sep 15-16  &    "      & 32 & B,V \\
1995 Sep 17-18  &    "      & 26 & B,V \\
1995 Oct 12-13  &    "      & 10 & B,V \\
1995 Oct 15-16  &    "      & 15 & V \\
1995 Oct 16-17  &    "      & 12 &  V \\
\hline
\end{tabular}
\end{table}

\subsection{Low-resolution spectroscopy}
A low-dispersion spectrum of Star~1 was obtained on the night
of 1991 October 7/8 using the {\it Carelec} spectrograph (Lemaitre et al. 1990) 
at the Cassegrain focus of the Haute-Provence 1.93--m telescope, 
equipped with a thinned, back-illuminated CCD (RCA No.1)
having $323\times512$ pixels of 30 $\mu$. A 5-min exposure was taken using
a 260 \AA/mm grating with a projected slit width of 2.4 arc--sec,
yielding an effective resolution of 15 \AA\ FWHM and covering the
wavelength range 3500-7300 \AA. Wavelength calibration was done using a
He--Ar lamp and flat-fielding using an internal Tungsten lamp. The
standard star Feige~25 was observed later in the night and was used to
derive a relative flux-calibrated spectrum for our candidate. All
reductions were made using the Spectroscopy package in MIDAS. 

\begin{figure*}
\epsfxsize=15cm
\centering{\hspace*{0.25cm}}
\epsfbox[56 448 564 782]{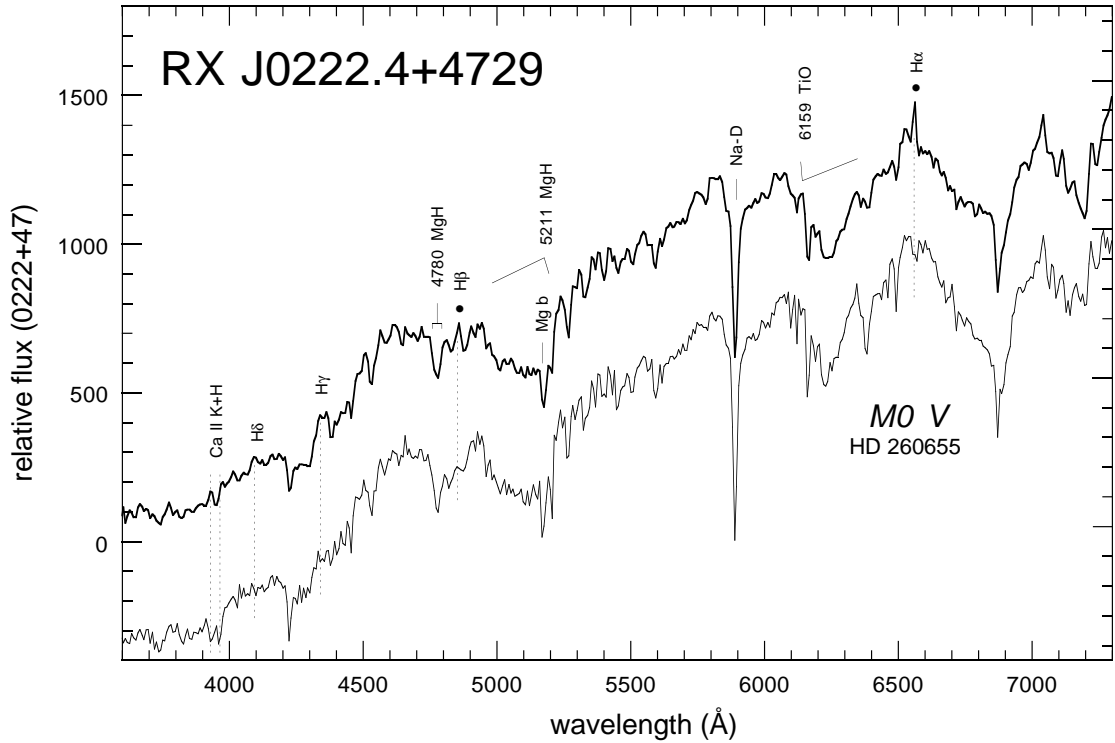}
\caption{The relative flux distribution of RX~J0222.4+4729 (Star~1) 
in the wavelength range $\lambda\lambda$ 3520-7340 \AA, obtained with 
{\it Carelec}, is shown together with the spectrum of the M0V star 
HD260655 (shown with a vertical offset)
extracted from the atlas of Jacoby et al. (1984). The continuum
and the absorption lines are similar in the two spectra with the noticeable
exception of the Balmer lines and the Ca~II H+K lines which appear in emission
in the spectrum of Star~1.}
\end{figure*}

The relative flux distribution of Star~1 in the wavelength interval $\lambda\lambda$
3520-7340 \AA\ is shown in Figure 2, together with the spectrum of the M0V
star HD~260655 extracted from the atlas of Jacoby et al. (1984). The
continuum and the absorption lines are almost identical in the two
spectra, which classifies Star 1 as a dwarf of spectral type M0.
Moreover, the presence of Balmer and Ca~II H+K emission lines 
is the signature of  chromospheric activity and strongly suggests that 
Star~1 is the optical counterpart of RX~J0222.4+4729. The optical variability
and binarity which we found (see later) further classify this 
object as a BY~Dra binary.

\subsection{Average Magnitude, Colours, Distance and X--Ray Luminosity}
CCD frames in the Cousins $B,V,R,I$\ bands were obtained during one
hour around the meridian on 1995 September~13, when seeing was poor due
to strong winds but weather conditions were photometric, useful for 
measuring the magnitudes of the bright stars of Figure 1 
with the exception of Star 5 which was measured on September~15 with 
better seeing but only through $B$ and $V$ filters. 

\begin{table}
\caption{Average magnitudes and colours 
for the stars in Figure~1 (1995~September)} 
\begin{tabular}[t]{ccccc}
\hline
Star No. & $V$   &   $B-V$   &  $V-R_{c}$ &   $V-I_{c}$ \\
\hline   
1 (0222.4+4729) & 11.10  & 1.44   &  0.89  &  1.74 \\
2              &  12.80  & 0.43   &  0.28  &  0.55 \\
3              &  13.60  & 0.41   &  0.28  &  0.54 \\
4              &  11.05  & 0.43   &  0.28  &  0.54 \\
5              &  12.96  & 0.94   &   & \\
6              &  11.65  & 0.72   &  0.42  &  0.79 \\
\hline
\end{tabular}
\end{table}

   The instrumental magnitudes were transformed to Cousins magnitudes
using previous calibrations of our equipment
(telescope+camera+filters) and using extinction coefficients
determined during nights of similar weather conditions. The resulting
uncertainty on the zero point is difficult to evaluate, but our Cousins
magnitudes and colours are probably accurate to a few percent. The
average $V$ magnitude and colours obtained in 1995 September for 
RX~J0222.4+4729 (Star 1) and for the comparison stars (2 to 6) are given
in Table 2. The inset frame at the bottom of Figure 1 was obtained with
the 0.8-m telescope (at f/15) and shows the surroundings of
Star 1 at twice the scale of Figure 1. The $V$ magnitude of the nearby star
at about 4.5 arc--seconds SE of the counterpart was measured using the
PSF-fitting routine NSTAR of DAOPHOT (Stetson 1987) on several frames
obtained on 1993 November~13 and is equal to $V=15.1$ (internal
dispersion $\pm$0.06 mag). This faint neighbour does not introduce errors
larger than a few mmag in the differential photometry of Star 1
through apertures of radius smaller than 3 arc--seconds. 
\begin{figure}

\epsfxsize=8.5cm
\centering{\hspace*{0.25cm}}
\epsfbox{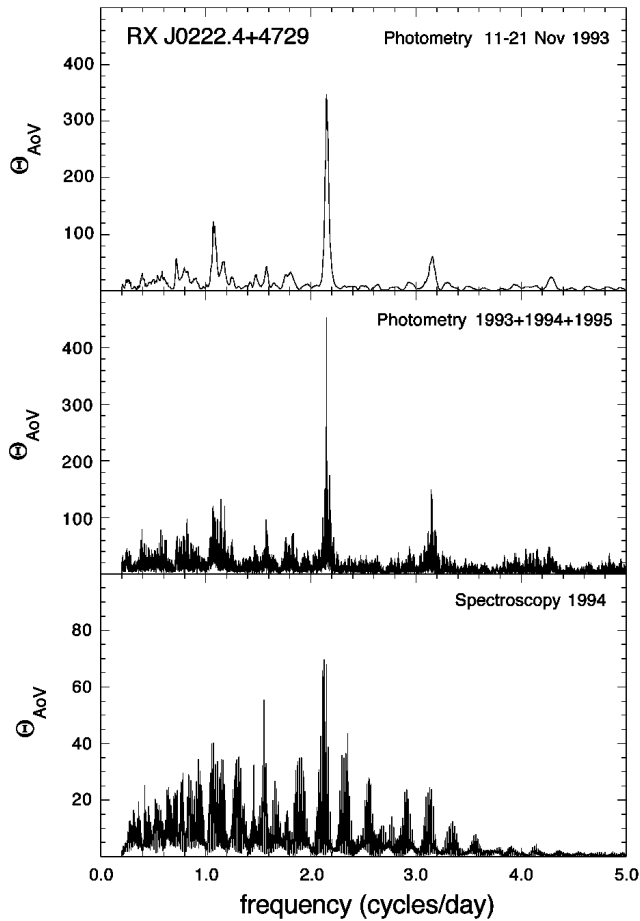}
\caption{AoV periodograms obtained by phase--binning 
of three data sets : {\bf Top (a)} Differential $V$--band 0.8--m CCD photometry (349 data points) 
obtained on 1993 November 11-21 (computed with a frequency 
step of 0.004 cycles/day). {\bf Middle (b)} The complete $V$--band photometry data set (615 
data points) obtained in 1993, 1994 and 1995 
(computed with a frequency step of 0.0005 cycles/day).
The 1993 data have been shifted
by 0.1 mag to compensate for the variations in average brightness between
1993 and 1994+1995.
{\bf Bottom (c)} Radial velocities of the metallic absorption lines 
(38 data points obtained in October and December 1994)
(computed with a frequency step of 0.001 cycles/day).}
\end{figure}

\begin{figure}
\centering{\hspace*{0.25cm}}
\epsfxsize=8.0cm
\epsfbox[52 426 544 792]{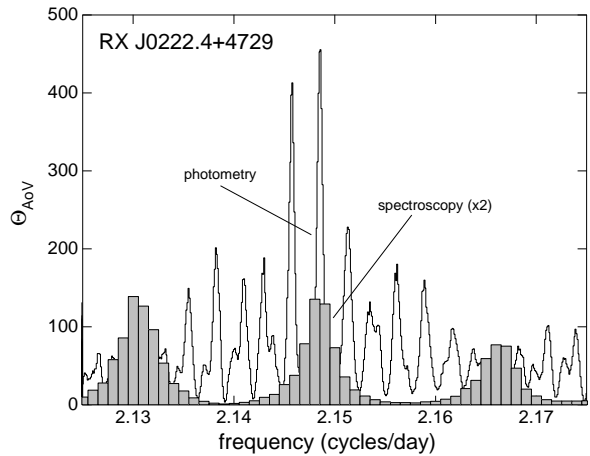}
\caption{The periodogram of Figure 3b has been recomputed with a higher frequency
resolution (0.0001 cycles/day) in the interval 2.125-2.175 cycles/day and is
plotted together with the corresponding part of the radial velocity periodogram 
of Figure 3c.}
\end{figure}

   Between 1993 November to 1995 October, the $V$ magnitude of 
Star~1 varied between 11.0 and 11.3 (see later the discussion on variability). 
Assuming $M_{v}$ in the range 8.5--9 (Allen 1973, Gray 1992), 
corresponding to the spectral type M0V, yields a 
distance in the range 25-35 parsecs for the optical counterpart of
RX~J0222.4+4729. The ROSAT counting  rate of $0.221 \pm 0.022$ c/s 
and hardness ratio HR1 = $0.06 \pm 0.09$
(Voges et al. 1996) give a 0.1-2.4 keV flux of 
$1.9\times10^{-12}$ erg cm$^{-2}$ s$^{-1}$, 
(using the conversion relation given by Fleming et al. 1995a), which yields
an X-ray luminosity of $1.9\times10^{29}$ erg s$^{-1}$ for a distance
of 30 pc, assuming negligible interstellar absorption. 
For $L_{bol} = 2.6\times10^{32}$ erg s$^{-1}$ (Allen 1973, Pettersen 1983)
we obtain an X-ray to bolometric luminosity ratio of $\log (L_{x}/L_{bol}) 
= -3.1 \pm 0.14$.
Results from the {\it Einstein}, EXOSAT and ROSAT soft X-ray surveys 
(Fleming et al. 1989, 1995b; Pallavicini et al. 1990; Schmitt et al. 1995) 
show that all K and M dwarf stars with Balmer emission
have X-ray luminosities in the range $10^{27}-10^{30}$ erg s$^{-1}$ and that for
M dwarfs $\log (L_{x}/L_{bol}) \simeq -3$, in agreement with the above
determination.

This BY~Dra binary close to the center of the error box 
is thus sufficient to account for the observed X-ray flux without having to 
consider the contribution from any other possible X-ray sources
in the error box.

\begin{figure*}
\epsfxsize=11cm
\centering{\hspace*{-1.0cm}}
\epsfbox[14 261 377 734]{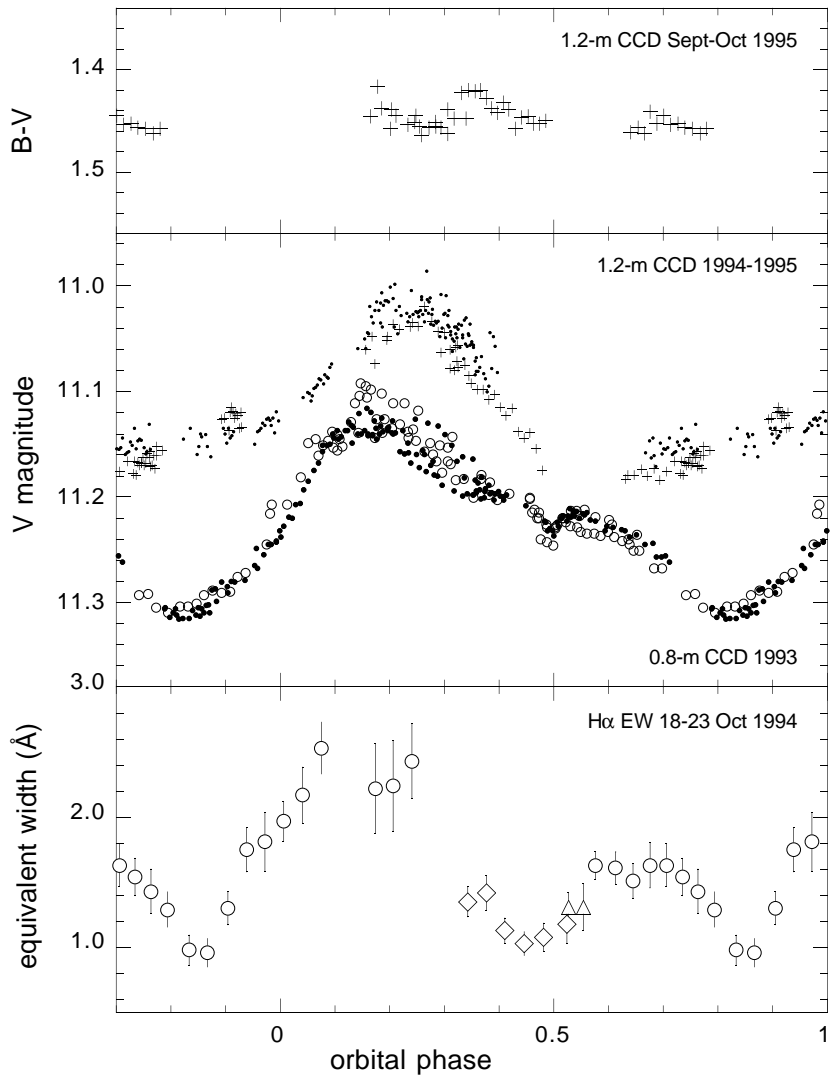}
\caption{{\bf Top (a)} The $B-V$ light curve obtained in 1995 (crosses) is plotted 
(crosses) as a function of the orbital phase (using
the same ephemeris as for Figure 6)
and shows no significant variation for a $V$ change
from 11.0 to 11.17.
{\bf Middle (b)} The $V$--band light curve obtained in 1993 (split in two
parts : 11-14 November 1993 (filled circles) 
and 16,19-21 November 1993 (open circles)), 1994 (dots)
 and 1995 (crosses) is plotted as a function of the orbital phase.
{\bf Bottom (c)} The equivalent width EW1 of the H$\alpha_{1}$ line (emitted by
the primary M0V star) is plotted as a function of the orbital phase (open
circles: 18 October, open triangles: 21 October and open diamonds: 23 October).}
\end{figure*}

\section{Photometric variability}
\subsection{Time analysis of photometric data}

   The 1993 November photometry was done differentially with respect
to  comparison stars 4 and 3 marked on Figure~1 and restricted to the
$V$ band. Our period search on this data set, containing 349 $\Delta V$
points, used the Analysis of Variance (AoV) method 
(Schwarzenberg-Czerny 1989) as implemented in the MIDAS Time-Analysis
package. This method is based on phase binning and does not make any prior
assumption concerning the shape of the light curve. The variance
statistic $\Theta$ was computed with a frequency step of 0.004 c/d and 6
bins. The resulting periodogram (Figure 3a) is unambiguous. A strong
($\Theta =350$) peak is present at a frequency close to 2.150 c/d,
corresponding to a period of 0.465 d, accompanied by smaller ones at
half the main frequency ($\Theta=120$), at 3.15 c/d  (a 1-day alias
with $\Theta=60$) and at twice the main frequency ($\Theta=30$).      
The photometric data acquired in 1994 November and December (185
$\Delta V$ data points) and in 1995 September and October (81
$\Delta V$ data points) were then added to the 1993 $\Delta V$ data
set with a shift of 0.1 mag to compensate for the variation in average
brightness of the source between 1993 and 1994--1995. The AoV
periodogram on the whole data set (615 $\Delta V$ data points) was
computed with a frequency step of 0.0005 c/d and 6 bins and is shown in
Figure 3b, confirming the peak of Figure 3a.

To refine the determination of
the peak frequency, we analyzed the frequency interval 2.125-2.175 c/d at
the higher resolution of 0.0001 c/d. The resulting AoV periodogram,
shown in Figure 4, exhibits a maximum $\Theta=450$ at a frequency 2.14855
c/d corresponding to a period $P=0.46543 \pm 0.00002$ d. This period, also
found in the radial velocity variations of Star~1 (which demonstrate the
binary character of the system --see next Section) is hence equal to the
orbital period.

\subsection{The light curve}

While the 1993 data obtained with the 0.8-m telescope yield complete phase
coverage, the 1994 and 1995 data sets, obtained with the 1.2-m telescope,
have unfortunate gaps between phases 0.4 and 0.6 and around phase 0.8 (see
Figure 5b). To better determine the shape of the 1993 light curve, we rejected
the data points of lowest S/N and in  some cases we averaged consecutive
data points with uncertainties $>$ 0.015 mag and separated by less than 5 minutes
in time. The resulting 1993 light curve of Figure 5b is plotted with filled
symbols for the nights of 11, 12, 13 and 14 November 1993, and with open
symbols for the five other nights (16, 17 and 19-21 November). The orbital
phase was computed from the ephemeris given in Section 4 below, phase 0.0
corresponding to the inferior conjunction of the primary M0V star and phase
0.5 to its superior conjunction.
Real changes of 0.02 to 0.05 mag can occur from one night to the next
in the light curve shape, mostly between phases 0.1 to 0.4 around maximum
light.

A minimum of depth 0.03-0.04 mag and lasting approximately 0.1 in
phase is visible at phase 0.5, indicating a partial eclipse and hence a high
inclination. With the exception of this shallow minimum, the light curve
(as in classical BY Dra stars) can be attributed to rotational modulation
of the visible star, its surface being unevenly darkened by starspots or
brightened by plages, the hemisphere observed at phase 0.75 being darker than
the opposite hemisphere observed at phase 0.25.

Figure 5b shows also that during the 1994 (dots) and 1995 observations (crosses),
the star was 0.1 magnitude brighter on the average. The shape of the light
curve was the same as in 1993 with the exception of the phase interval 0.6-1.0.
where the minimum was filled in. Figure 5a shows the relative constancy of
the $B-V$ color index observed in 1995 September and October.

\section{High-resolution spectroscopy} 
\subsection{Observations}

38 high-resolution spectra were obtained on several nights in 1994 October
and December using the {\it Elodie} spectrograph (Baranne et al.
1996) at the Haute-Provence 1.93--m telescope (see Table 3 for the
observing log). {\it Elodie} is a
cross-dispersed echelle spectrograph located in a
temperature-controlled room and fed by optical fibers from the
Cassegrain focus. A single exposure gives a spectrum, covering the
interval 3906 \AA\ to 6811 \AA\  with a resolution of $\cal{R}
\simeq$ 42000, distributed over
67 equally-spaced orders, which is recorded on a thinned,
back-illuminated $1024\times1024$ Tektronix CCD (TK1024 No.2). 
Two entrance apertures, one
for the star and one for the sky, spanning each 2 arc--sec on the sky and
separated by 1.8 arc--min, yield two interleaved sets of 67 orders each. 

Wavelength calibration was done using a Thorium lamp and flat-fielding
using a tungsten lamp, both installed in the Cassegrain adapter. The
entire reduction procedure, including flat-fielding,
cosmic-ray rejection, wavelength calibration and optimal extraction, 
is automatic and is done, immediately after the
image is read out, by a pipeline software process. The observer can
proceed a step further by cross-correlating the reduced spectrum,
corrected to the barycenter of the solar system, with different
software masks derived from Coravel-type hardware masks (software "slots"
described by two edge wavelengths for each selected line). Two such
masks are normally available, F0 and K0, useful for spectral
types from F0 through M0. Only metallic absorption lines are
included in the masks and the radial velocities measured are those of the mean
of the lines present in the spectra, normally numbering around 3000.
Given the late-type spectrum of our star we used the K0 mask for all
on-line reductions.

The correlation profile obtained is then displayed 
and an interactive fitting procedure
featuring one (or two) gaussian profile(s) yields the radial velocity for
the object(s) just observed. The actual uncertainty for the measured
radial velocity depends on the width and depth of the cross-correlation
function (CCF), and on the S/N achieved (see Queloz 1995
and Baranne et al. 1996). In a few cases where spectra were
taken during bright moon conditions, when the correlation profile was
affected by scattered moonlight, a correction was applied using
the correlation profile derived from the sky spectrum. 

The spectra of RX~J0222.4+4729 were taken using exposure times ranging
from 1000 to 2000 seconds, with a mean S/N ratio of 10 (at 5500 \AA),
but which ranged actually from 5.7 through 16, due to changes in
seeing and atmospheric transparency. For our observations of 
RX~J0222.4+4729, the expected error for a velocity determined 
from a S/N=10 spectrum is around 1 km/s. 
An experimental determination of the actual errors
is given below.

While the on-line correlation process extracts the velocity information from the 
 metallic absorption lines, individual line profiles for particular features 
can be examined off-line  in the individual spectra. The behavior of the H$\alpha$
emission profiles is reported on below. 
The $\lambda 6708$ \AA\ Li~I absorption line is
not detected in our best signal-to-noise ratio {\it Elodie} spectra although 
it lies in the wavelength region where both the CCD and the spectrograph have a
high efficiency. Also, although detected in emission on our low-resolution 
{\it Carelec} spectra (Figure~2), the Ca~II H and K 
lines are not measurable with {\it Elodie},
even though they lie in the first few recorded orders. This is because
the efficiency of both the CCD and the spectrograph system (including the
optical fibers) are very low at these wavelengths. Much longer exposures
would have been needed to achieve detection.

\begin{figure}[h]
\epsfxsize=8.7cm
\epsfbox[5 431 568 837]{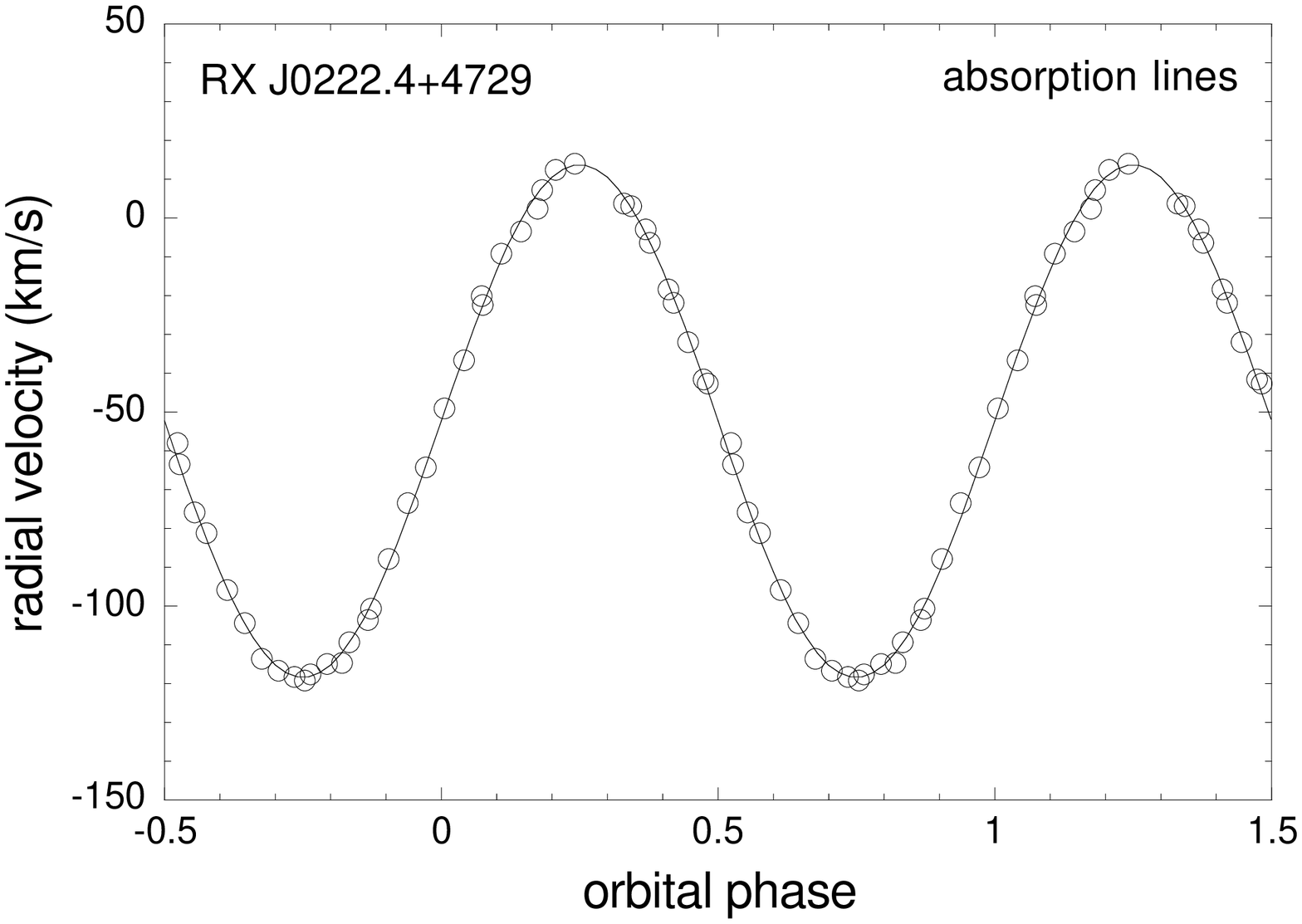}
\caption{The 38 radial velocity measurements for the metallic absorption 
lines of the M0V star in RX~J0222.4+4729, obtained with {\it Elodie}, 
are plotted {\it vs.} the orbital phase 
computed from the ephemeris given in the text.
$\phi = 0$ corresponds to an inferior conjunction of the M0V primary. The
solid line represents a least--squares fit to a sine curve.}
\end{figure}

\subsection{Radial velocity of the absorption lines}

For each of the 38 spectra obtained with {\it Elodie}, the average radial
velocity of the absorption lines was derived from the
cross-correlation function (CCF) of the observed spectrum with a K0
mask. A single broad Gaussian (implying rapid rotation --see later) was 
always an acceptable fit to the CCF. The variability which appears in the
radial velocity values listed in Table~3 clearly shows that Star~1 belongs
to a binary system.
The period search on this data set  used also the AoV statistic,
with 7 bins and a frequency step of 0.001 c/d. The periodogram is
shown in Figure 3c. The two epochs of observation separated 
by two months (1994 October and 1994 December) resulted in 3 equally
probable peaks in the frequency interval 2.10-2.15 c/d. One of them
coincides with the photometric peak as is shown in Figure 4 which is a
close-up view of both photometric and radial velocity periodograms in
the frequency interval 2.125-2.175 c/d. By attributing the
photometric variability to rotational modulation 
of the M0V star, we conclude that the star's rotation is
synchronized with its orbital motion, a result expected for an orbital
period as short as 0.465 d.

   We made a least-squares fit to obtain the best match of the 
radial velocities to a sine curve, starting with an initial value
of $P=0.465$ d. The four-parameter fit (period, phase, semi-amplitude
and zero point), computed using the MIDAS Time-Series Analysis package, 
yielded a period $P=0.46543 \pm 0.00001$ d (11.1703 hours),
an epoch $T_{\circ}$ = HJD$\ 2449644.1042 \pm 0.0014$ for phase 0.0,
where $T_{\circ}$ is defined as the time of an inferior conjunction of the M0V
star (preceding the time of maximum redshifted velocity by 0.25 cycle),
a semi-amplitude $K_{1}=66.0 \pm 0.6$ km/s and a systemic velocity
$\gamma =-52.4 \pm 0.3$ km/s. The radial velocity curve of RX~J0222.4+4729 is
shown in Figure~6 where the radial velocities determined from the
absorption lines of the M0V star spectra are plotted as open circles
and the fit to a sine curve as a solid line. The goodness of the fit
is an evidence for a circular orbit. The individual radial velocities
are listed in Table~3 where we also give the S/N (at 5500 \AA) 
of the spectra and the velocity residuals to the sine fit
(in km/s). The standard deviation of the latter is 1.1 km/s which is
close to the expected uncertainty mentioned above.

   The average FWHM of the absorption-line CCF yields an estimate
of the projected rotational velocity of the M0V star: $v_{rot} \sin i =
2 \pi R_{1} \sin i /P \simeq 85 \pm 5$ km/s, where $R_{1}$ is the M0V 
star radius.

\begin{table}
\renewcommand{\footnoterule}{\rule{5mm}{0mm}\vspace{-2mm}}
\caption[]{Absorption-line radial velocity data for RX~J0222.4+4729} 
\begin{flushleft}
\begin{minipage}{8.7cm}
\vspace{-\abovedisplayskip}
\begin{tabular}[8.7cm]{crrrr}
\hline
HJD\footnote{HJD--2440000 at midpoint of exposure} &  
$V_{rad}\footnote{M0V star radial velocity of metallic
absorption lines}$ & $\epsilon\footnote{residuals from a sine fit}$ & 
$\phi_{orb}$\footnote{using $P=0.46543$ d and
 $T_{\circ}=9644.1042$} & S/N\footnote{at 5500 \AA}\\
(d) & (km/s) & (km/s)& &\\
\hline
9644.3721 & --81.13 & +1.40  & 0.576 & 13.2\\
9644.3896 & --95.79 & --0.31  & 0.613 & 11.4\\
9644.4043 & --104.40 & +0.06  & 0.645 &9.9\\
9644.4189 & --113.60 & --2.19  & 0.676 & 8.7\\
9644.4326 & --116.60 & --0.77  & 0.706 & 9.0\\
9644.4463 & --118.30 & --0.21  & 0.735 & 9.9\\
9644.4600 & --117.50 & +0.62  & 0.764 & 7.6\\
9644.4736 & --114.80 & +1.12  & 0.794 & 8.6\\
9644.4922 & --109.30 & +0.21  & 0.834 & 7.7\\
9644.5078 & --103.60 & --2.28  & 0.867 & 7.9\\
9644.5254 & --87.80 & +1.75  & 0.905 & 9.3\\
9644.5410 & --73.44 & +3.87  & 0.939 & 9.0\\
9644.5566 & --64.25 & --0.29  & 0.972 & 7.2\\
9644.5723 & --48.98 & +1.13  & 0.006 & 11.8\\
9644.5889 & --36.58 & --1.07  & 0.041 & 9.1\\
9644.6045 & --22.41 & +0.13  & 0.075 & 11.5\\
9644.6504 &  +2.52 & --3.55  & 0.174 & 5.9\\
9644.6660 & +12.38 & +1.22  & 0.207 & 5.7\\
9644.6816 & +14.08 & +0.65  & 0.241 & 7.7\\
9647.6074 & --63.38 & +0.09  & 0.527 & 10.9\\
9647.6201 & --75.80 & --1.40  & 0.554 & 6.6\\
9649.3838 &  +3.07 & +0.56  & 0.343 & 10.5\\
9649.3994 &  --6.43 & --0.07  & 0.377 & 9.6\\
9649.4150 & --18.32 & --1.06  & 0.411 & 10.2\\
9649.4316 & --32.01 & --1.46  & 0.446 & 10.4\\
9649.4482 & --42.68 & +2.25  & 0.482 & 9.3\\
9649.4678 & --58.02 & +4.25 & 0.524 & 7.2\\
9699.3760 & --119.10 & --0.74  & 0.754 & 16.4\\
9699.5244 & --20.07 & +3.00  & 0.073 & 15.6\\
9699.5410 &  --9.19 & +1.49  & 0.109 & 13.7\\
9699.5576 &  --3.48 & --3.11  & 0.144 & 13.6\\
9699.5752 &  +7.22 & --0.49  & 0.182 & 11.7\\
9703.3672 &  +3.73 & --1.70  & 0.330 & 9.6\\
9703.3857 &  --2.86 & +1.38  & 0.369 & 9.9\\
9703.4092 & --21.82 & --1.18  & 0.420 & 11.8\\
9703.4346 & --41.56 & +0.36  & 0.474 & 13.9\\
9703.5957 & --114.60 & --2.63  & 0.821 & 12.2\\
9703.6201 & --100.60 & --1.02  & 0.873 & 10.9\\
\hline
\end{tabular}
\end{minipage}
\end{flushleft}
\end{table}

\subsection{The H$\alpha$ emission lines}
\subsubsection{Radial velocity}

   All the {\it Elodie} high-resolution spectra showed strong H$\alpha$ emission
above the continuum. Figure 7 illustrates the profile at representative
phases along an orbital cycle. At inferior (phase 0) and superior (phase 0.5)
conjunctions of the primary M0V star, the H$\alpha$ line showed an
emission  profile well matched to a single Gaussian profile, while a
second component was present at intermediate phases, appearing
blueward of the primary component between phases 0.1 and 0.4 and
redward between phases 0.6 and 0.9. When both components were present,
a bi-Gaussian fit, done off-line with the MIDAS Fitting package,
 yielded the central wavelengths of both components
H$\alpha_{1}$ and H$\alpha_{2}$, as well as the equivalent width of 
the primary component above the local continuum. 
Given the strength of the emission, we have not corrected the
H$\alpha$ equivalent width for the underlying photospheric absorption
(such a correction would add a constant term of a few tenths of an \AA).

   The resulting heliocentric radial velocities of H$\alpha_{1}$ and 
H$\alpha_{2}$ are
plotted in Figure 8 as a function of the orbital phase 
derived from the primary absorption lines. A least-squares fit of the
H$\alpha_{1}$ radial velocities to a sine curve yields the same parameters,
within the uncertainties, as those derived from the absorption lines:
$P=0.46540 \pm 0.00008$~d, $T_{\circ} =$ HJD $2449644.103 \pm 0.004$, $K_{1}
= 66.2 \pm 1.9$ km/s, which confirms that the  H$\alpha$ emission line
of the primary has the same radial velocity curve as its photospheric
absorption lines. Figure 8 illustrates this result, since the dotted line
plotted on the H$\alpha_{1}$ radial velocities, which is the same
sine curve plotted in Figure 6 on the absorption-line radial velocities,
matches the data very well. 
  
\begin{figure}
\epsfxsize=9.0cm
\epsfbox{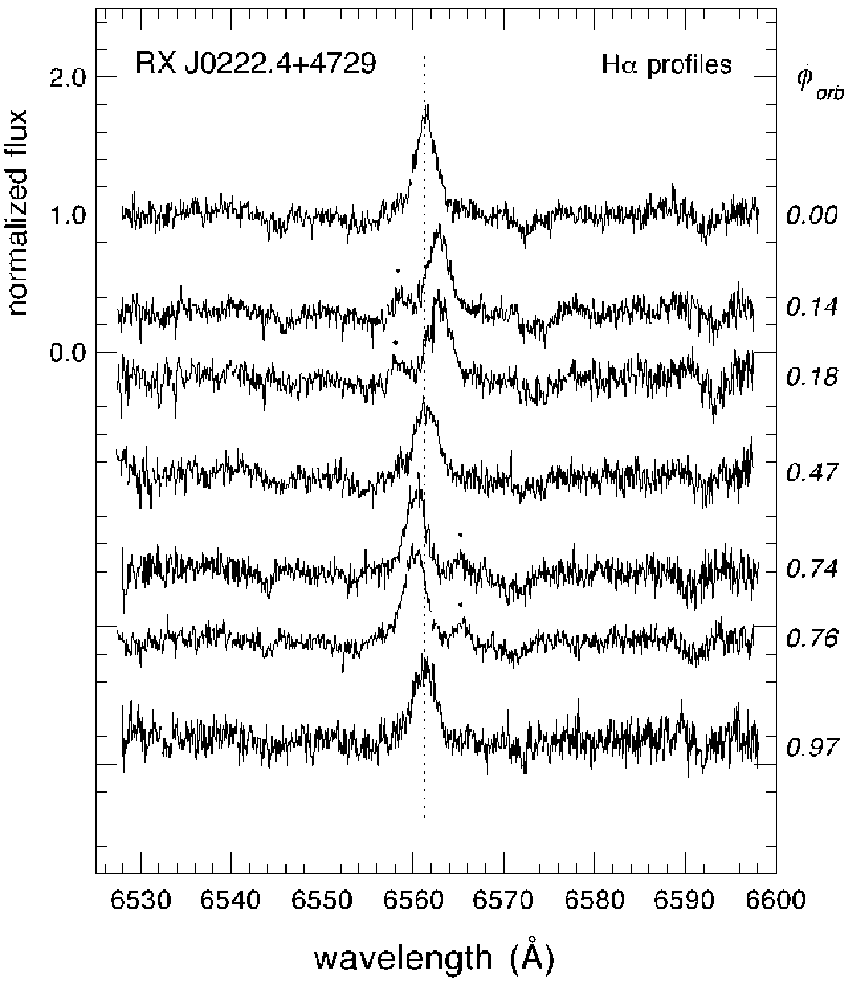}
\caption{Individual {\it Elodie} spectra around  H$\alpha$ are 
shown near conjunction (phases 0.0 and 0.5) and quadratures 
(phases 0.25 and 0.75). The  H$\alpha$ line emitted by the secondary
 star appears blueward of the primary component near
phase 0.2 and redward near phase 0.7 while the profile appears 
single-Gaussian near phases 0.0 and 0.5. The dotted vertical line is 
located at the systemic velocity.}
\end{figure}

\begin{figure}
\epsfxsize=8.7cm
\epsfbox[10 448 541 819]{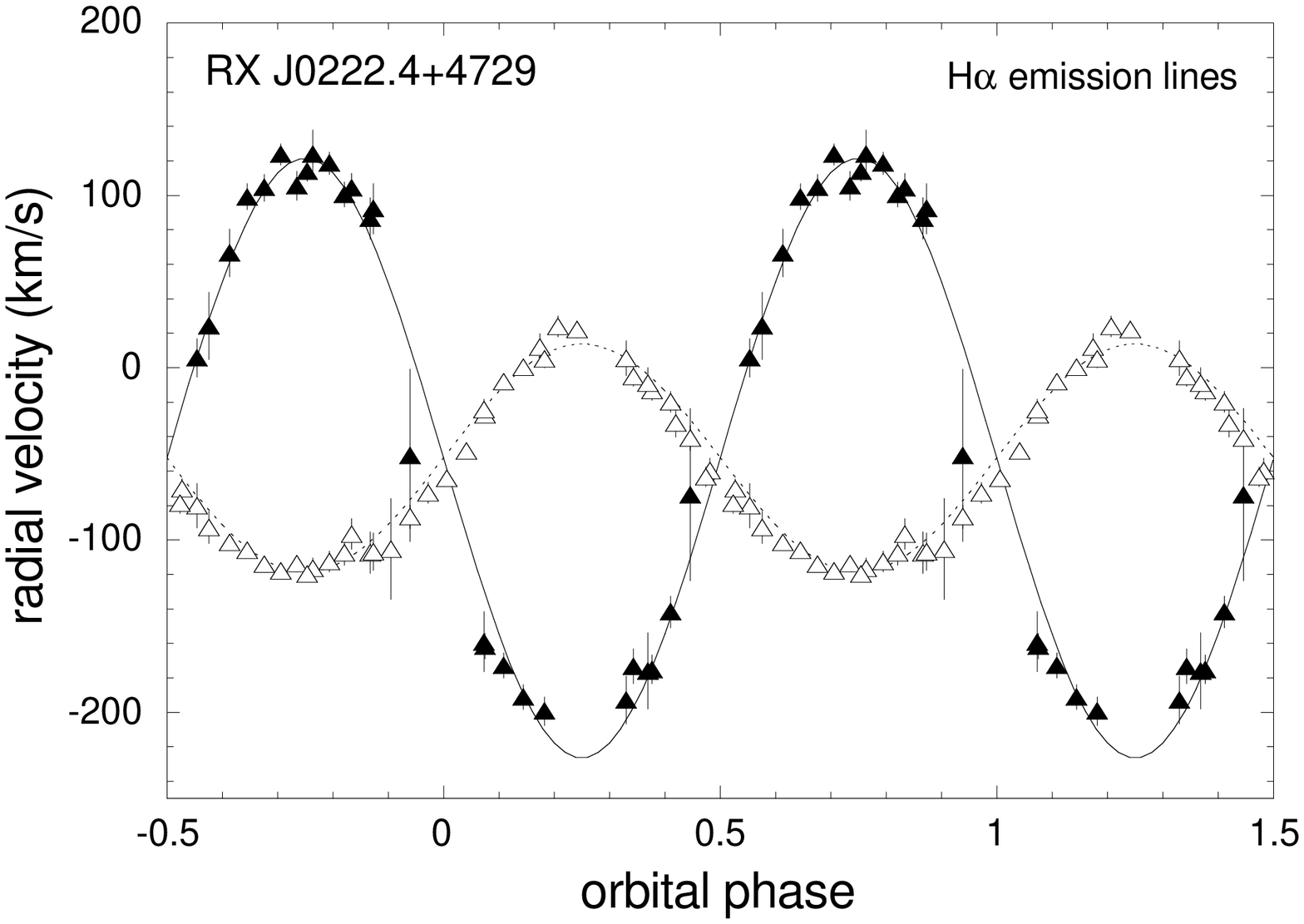}
\caption{
The {\it Elodie} radial velocities of the H$\alpha$ lines emitted by the primary M0V star
(open triangles) and the secondary star (filled triangles) are plotted as a function
of the orbital phase (computed with the same ephemeris as in Figure 5). The
dotted line curve overplotted on the triangles (primary star) is the fit
derived for the metallic absorption lines of the primary and shown
in Figure 5.}
\end{figure}

The fit of the H$\alpha_{2}$ radial velocities to a sine  curve yields
 $P = 0.46547 \pm 0.00008$~d, $T_{\circ} =$ HJD $2449644.33 \pm 0.01$, $K_{2}=174
\pm 5$ km/s, adopting for H$\alpha_{2}$ the same systemic velocity as for the
other lines, and is plotted  in Figure 8 as a solid line on the H$\alpha_{2}$
radial velocity data (filled triangles). The radial velocity of H$\alpha_{2}$
varies with the same period as the radial velocities of H$\alpha_{1}$ and 
of the absorption lines but shows a phase shift of $180^{\circ}$ corresponding
to the difference of $P/2$ in $T_{\circ}$. We can thus assume that the
radial velocities of the emission  lines H$\alpha_{1}$ and H$\alpha_{2}$ reflect the
orbital motion of the primary and secondary stars respectively.
Although too faint to contribute to the continuum or to the absorption line
spectrum, the secondary star reveals itself by the strong
intensity of its H$\alpha$ emission. 

Assuming a mass ratio $q=M_{2}/M_{1}$ equal to the ratio of amplitudes  
$K_{1}/K_{2}$ yields $q=0.38 \pm 0.04$. A reasonable estimate
for the mass of the primary star is  $M_{1}=0.52 M_{\odot}$ (Gray
1992), with an uncertainty of at least 10\%. From the mass ratio $q$ we
derive for the secondary star 
$M_{2} \sim 0.2 M_{\odot}$ (with a 20\% uncertainty) which corresponds
to  a dwarf of spectral type M5, and $M_{1} + M_{2}=(0.72 \pm 0.07)
M_{\odot}$

Since the primary star is more than 3 magnitudes
brighter than the secondary in the Cousins {\it R} band, the continuum
intensity of its spectrum around H$\alpha$ should be more than 10   
times brighter than the continuum intensity of the secondary and yet
the H$\alpha$ emission of the secondary appears above the primary's
continuum at orbital phases near quadratures. We may safely conclude
that the chromospheric activity of the secondary star exceeds that,
already quite intense, of the primary. 

\subsubsection{Equivalent widths}

As mentioned above, the H$\alpha$ lines appear in pure emission with Gaussian
profiles, indicating that the underlying photospheric absorption is relatively
small. Excluding conjunction, the equivalent widths EW1 and EW2 were
measured by fitting a Gaussian to each profile and a polynomial to the continuum
(emitted by the primary). EW1 was found to vary between 1 and 3 \AA\ while
EW2 rarely exceeded 0.1 \AA\, and we have neglected the contribution of EW2
to EW1 near conjunction since this contamination is small compared to the
uncertainty on the placement of the local continuum.

In Figure 5c the equivalent width EW1 measured on the nights of 1994 October
18, 21 and 23 is plotted as a function of orbital phase. The EW1 modulation
appears directly correlated to the 1993 $V$-band light curve obtained 11 months
earlier (unfortunately the $V$ light curve obtained in November and December
1994 shows too many phase gaps to be useful). Maximum EW1 occurs between
phases 0.1 and 0.3 as the maximum in $V$; similarly a deep minimum occurs near
phase 0.8 in both curves. The H$\alpha_{1}$ variability is apparently dominated by
rotational modulation and not by flares.  

Our results differ from most studies which indicate an anti-correlation of
these activity tracers in BY~Dra and RS~Cvn stars. Such an
anti-correlation is not a general rule however. Other instances of
correlation are known, for example one has been observed in the single
M dwarf Gliese~890 (HK~Aqr) of the BY~Dra type by Young et al. (1990),
another one in the short-period eclipsing RS~Cvn binary BH~Vir (L\'azaro
\& Ar\'evalo 1994). 

The main difference between the EW1 modulation  and the $V$ light curve
is observed near the primary's superior conjunction. A deep and broad minimum
in the EW1 modulation, centered near phase 0.45, corresponds to the shallow
and narrow mininum at phase 0.5 in the $V$ light curve which indicates a partial
eclipse of the primary's disk. This suggests that part of the H$\alpha_{1}$
emission comes from co-rotating prominence-like material surrounding the
primary which may be viewed off the limb, the eclipse of the prominence by
the secondary occuring earlier in phase than the partial eclipse of the primary's
disk.
                          
\section{Discussion}  

The results presented in this paper yield the  orbital elements
summarized in Table 4.

\begin{table}
\caption{Orbital elements for RX~J0222.4+4729} 
\begin{tabular}{r@{\ =\ }l}

$K_{1}$ & $66.0 \pm 0.6$ km/s  \\
              
$K_{2}$ & $174 \pm 5$ km/s  \\
              
$a \sin i$ & $(2.189 \pm 0.064) R_{\odot}$ \\
   
$\gamma$ & $- 52.4 \pm 0.3$ km/s \\
               
$(M_{1}+M_{2})\ {\sin}^{3}i $ & $(0.666 \pm 0.058) M_{\odot}$ \\
       
$q \equiv M_{2}/M_{1} $ & $0.38 \pm 0.04$ \\
     
$P$ & $0.46543 \pm 0.00001$ d \\
      
$T_{\circ}$ & HJD $2449644.1042 \pm 0.0014$ \\
            
$e$ & 0.0     \\   
\end{tabular}
\end{table}

The rotational
velocity of the primary star has been derived from the average FWHM of
the cross-correlation function of its absorption lines with a K0 mask giving 
$v_{rot} \sin i = 2 \pi R_{1} \sin i /P \simeq
85 \pm 5$ km/s, where $R_{1}$ is the radius of the primary star. 
From this we derive 
$R_{1}/a = 0.35 \pm 0.03$  and  $R_{1} = 0.78 \pm 0.09 R_{\odot}$.  
This radius is somewhat larger than the average estimate for a main
sequence M0V star but is intermediate between the radii adopted for
BY~Dra (M0V) and for CC~Eri (K7V) (Strassmeier et al. 1993) 
which are both active binaries of the same type. 

The grazing-incidence angle between the two stars is given by $\tan i_{graz}
= 2.19/(R_{1}+R_{2})$, $R_{1}$ and ${R}_{2}$  being the radii of the primary
and secondary stars respectively. Assuming for $R_{1}$ the value given above
and for $R_{2}$ a value compatible with that for $M_{2}$ ($\sim 0.25$-$0.30 R_{\odot}$)
gives $R_{1}+R_{2}$ in the range 1.0-1.1 $R_{\odot}$ and $i_{graz} \sim 64^{\circ}$.
On the other hand, combining the total mass derived earlier, $M_{1} + M_{2}=(0.72 \pm 0.07)
M_{\odot}$, with the value of $(M_{1}+M_{2})\ {\sin}^{3}i $ 
yields an orbital inclination in the
$66$--$90^{\circ}$ range, with a central value of  $i \sim 77^{\circ}$, thus
partial eclipses  must occur.

At phase 0 the eclipse of the secondary is not detectable in the $V$ light
curve due to its faintness, while the minimum observed at phase
0.5 is compatible with a partial eclipse of the primary.  Its shape, small
depth and duration could agree with an inclination in the range $70-80^{\circ}$.
The eclipse profile may be altered by starspot groups and plages on the primary
and eventually by flaring on both stars.

Considering the strength of the Balmer emission and the amplitude of the
$V$ light curve, this system is one of the most active BY~Dra binaries. Variations
of both the H$\alpha$ emission and the $V$ light curve are clearly dominated
by rotational modulation and not by flaring. The $V$ light curve and the H$\alpha$
equivalent width of the primary are directly correlated, which is unusual
among BY~Dra binaries. This correlation and the minimum observed 
in the H$\alpha$ equivalent width near the
primary's partial eclipse suggest that Balmer emission occurs in large emitting loops
connecting the major spot groups. The coronal X-ray emission is also a good
tracer of stellar activity. For RX~J0222.4+4729, a rapid rotator ($v \sin
i \sim 85$ km/s), we find an X-ray to bolometric
luminosity ratio of $\log (L_{x}/L_{bol}) \sim -3.1 \pm 0.14$. This ratio
is in excellent agreement with the dependence  of $\log (L_{x}/L_{bol})$
on projected rotational velocity derived by Stauffer et al. (1994) for Pleiades
members, which shows that for $v \sin i > 15-20$ km/s, $\log (L_{x}/L_{bol})$
is constant and equal to $-3$. This supports the concept of saturation of coronal
X-ray emission for the most rapidly rotating late-type stars, in particular
for the BY~Dra binaries (Pallavicini et al. 1990).

\acknowledgements
{We are indebted to the Max-Planck Institut f\"ur Extraterrestrische Physik
and to Ch.~Motch in particular for communicating us the 
preliminary ROSAT source position. We thank the OHP night assistants for 
their participation in the observations
at the 0.8-m and 1.2-m telescopes. The 1.93--m {\it Elodie} spectrograph 
and the 1.2-m CCD camera system were funded in part 
by the PACA Regional Council.  The Digitized Sky Survey was produced at the
Space Telescope Science Institute (STScI) under U.S. Government Grant NAG
W-2166.}


\begin{thebibliography}{}

\bibitem{}Allen C.W. 1973, Astrophysical Quantities, Athlone Press, London

\bibitem{}Baranne D., Queloz D., Mayor M., Adrianzyk G., Knispel G., Kohler D.,
Lacroix D., Meunier J.P., Rimbaud G., Vin A. 1996, A\&AS 119, 373

\bibitem{}Fleming T.A., Gioia I.M., Maccacaro T. 1989, ApJ 340, 1011

\bibitem{}Fleming T.A., Molendi S., Maccacaro T., Wolter A. 1995a, ApJSS 99, 701

\bibitem{}Fleming T.A., Schmitt J.H.M.M., Giampapa M.S. 1995b, ApJ 450, 401

\bibitem{}Gray D.F. 1992, The Observation and Analysis of Stellar Photospheres,
Cambridge University Press, Cambridge

\bibitem{}Howell S.B. 1989, PASP 101, 616

\bibitem{}Jacoby G.H., Hunter D.A., Christian C.A. 1984, ApJS 56, 257

\bibitem{}L\'azaro C., Ar\'evalo M.J. 1994, Cool Stars, Stellar Systems and the
Sun (8th Cambridge Workshop), J.-P.~Caillault (Ed.), 
ASP Conf. Series, Vol.64, p. 436

\bibitem{}Lemaitre G., Kohler D., Meunier J.P., Vin A. 1990, A\&A 228, 546

\bibitem{}Pallavicini R., Tagliaferri G., Stella L. 1990, A\&A 228, 403

\bibitem{}Pettersen B. 1983, in Activity in Red-dwarf stars, P.B. Byrne,
M. Rodon\`o, Eds., Reidel, Dordrecht, p.17

\bibitem{}Queloz D. 1995, IAU Symposium No. 167, New Developments in Array
Technology and Applications, A.G. Davis Philip et al., Eds., Reidel, Dordrecht,
p.221

\bibitem{}Schmitt J.H.M.M., Fleming, T.A. Giampapa M.S. 1995, ApJ 450, 392

\bibitem{}Schwarzenberg-Czerny A. 1989, MNRAS 241, 153

\bibitem{}Stauffer J.R., Caillault J.P., Gagn\'e M. et al. 1994, ApJS 91,
625

\bibitem{}Stetson, P.B. 1987, PASP 99, 101

\bibitem{}Stetson, P.B. 1990, PASP 102, 932

\bibitem{}Strassmeier K.G., Hall D.S., Fekel F.C., Scheck M. 1993, A\&AS 100, 173

\bibitem{}V\'eron-Cetty M.P., V\'eron, P. 1996, A\&AS 115, 97 

\bibitem{}Voges W., Aschenbach B., Boller Th. et al. 1996, A\&AS in press

\bibitem{}Young A., Skumanich A., MacGregor K.B., Temple S. 1990, ApJ 349, 608


\end{thebibliography}
\end{document}